# Effect of cluster size of chalcogenide glass nanocolloidal solutions on the surface morphology of spin-coated amorphous films

T. Kohoutek[a)], T. Wagner and M. Frumar
*Department of Inorganic Chemistry, Faculty of Chemical-Technology, University of Pardubice, Legion´s sq. 565, 53210, Czech Republic*

A. Chrissanthopoulos
*Foundation for Research and Technology Hellas, Institute of Chemical Engineering and High Temperature Chemical Processes, FORTH/ICE-HT, P.O. Box 1414, GR-26 504 Patras, Greece and Department of Materials Science, University of Patras, GR-26 504 Patras, Greece*

O. Kostadinova and S. N. Yannopoulos
*Foundation for Research and Technology Hellas, Institute of Chemical Engineering and High Temperature Chemical Processes, FORTH/ICE-HT, P.O. Box 1414, GR-26 504 Patras, Greece*

**Abstract**
Amorphous chalcogenide thin film deposition can be achieved by a spin-coating technique from proper solutions of the corresponding glass. Films produced in this way exhibit certain grain texture, which is presumably related to the cluster size in solution. This paper deals with the search of such a correlation between grain size of surface morphology of as-deposited spin-coated $As_{33}S_{67}$ chalcogenide thin films and cluster size of the glass in butylamine solutions. Optical absorption spectroscopy and dynamic light scattering were employed to study optical properties and cluster size distributions in the solutions at various glass concentrations. Atomic force microscopy is used to study the surface morphology of the surface of as-deposited and thermally stabilized spin-coated films. Dynamic light scattering revealed a concentration dependence of cluster size in solution. Spectral-dependence dynamic light scattering studies showed an interesting athermal photo-aggregation effect in the liquid state.

---
[a)] Electronic address: tomas.kohoutek@upce.cz
[b)] Electronic address: sny@iceht.forth.gr

# I. INTRODUCTION

The capability of fabricating and tailoring in a consistent way thin films of amorphous semiconductors is essential for modern electronic devices. Thin films structure and hence their properties that are important for applications depend strongly upon the deposition method. Methods that are typically employed for thin film deposition include thermal evaporation, magnetron sputtering, pulsed laser deposition, etc. Main shortcomings are encountered with the above techniques ranging from poor control of materials compositions (for evaporation techniques) or high cost of film fabrication. On the other hand, wet-chemistry methods, where film deposition is achieved through spin coating of material's solution, have recently revived in view of their simplicity, reliability in depositing thin (and ultrathin) films and low cost of operation. Fabrication of metal chalcogenides thin films of technological importance has recently been demonstrated [1, 2].

Arsenic chalcogenide As-X (X: S, Se) bulk glasses and their amorphous thin films have been systematically investigated over the last decades [3-5] owing to the wide range of applications that these materials meet. Photo-sensitivity is the key issue that can be exploited to alter and control material's structural details which ultimately determine other macroscopic properties such as mechanical, optical (linear and nonlinear), thermodynamic, rheological, etc. [4, 5]. Obviously, thin films are more exploitable in the application sector (e.g. as planar waveguides) than bulk glasses and hence have attracted most of the scientific interest. This is evidenced by the fact that various routes have been employed to fabricate thin films, as mentioned above, which eventually lead to different local atomic arrangements and hence to materials with different properties. Indeed, it is now well documented that arsenic chalcogenide thin films produced by vacuum deposition techniques are characterized by different structural details.

The fact that thin films of amorphous chalcogenides can be prepared by means of the spin-coating method from solutions of chalcogenide glasses has been demonstrated long ago [7]. Arsenic sulphide glasses can readily be dissolved in organic solvents such as amines over the concentrations range 1-$10^3$ mg ml$^{-1}$. Chalcogenide spin-coated films have demonstrated similar quality and properties [8] to those films deposited using the vacuum techniques although subtle differences between them were also reported [7(d)].

Attempts to understand the mechanism of dissolution of chalcogenides in amines have been reported [7] and the existence of the presence of chalcogenide clusters with dimensions of several nanometers was envisaged, see Fig. 1. Further studies were focused on the surface morphology of as-prepared spin-coated As-S chalcogenide thin films deposited from amine solutions. In particular, a nanoscale grainy character of such film surfaces was revealed using transmission electron microscopy (TEM) [7(d)] and atomic force microscopy (AFM) techniques [9]. The film surface morphology of as-deposited spin-coated films is probably influenced by solvent evaporation during film fabrication while film's thickness depends mainly on spinning parameters [10], solution viscosity and adhesion of chalcogenide material to the substrate. There are several factors influencing the resulting quality of surface morphology of spin-coated films which are crucial for their applications, e.g. as high resolution photoresists. Factors such as the use of suitable solvent for bulk glass dissolution (solubility, solution viscosity, etc.), film preparation conditions (inert atmosphere), adhesion to substrate, spin speed and time, film stabilization have to be particularly considered in order to reach the desired film quality. Appreciating that film structure is usually fragmented due to the existence of large grains – an undesired outcome for applications – it would be important to ask if there is an particular relation between cluster size in the solution of the chalcogenide glass and grain size in the film morphology.

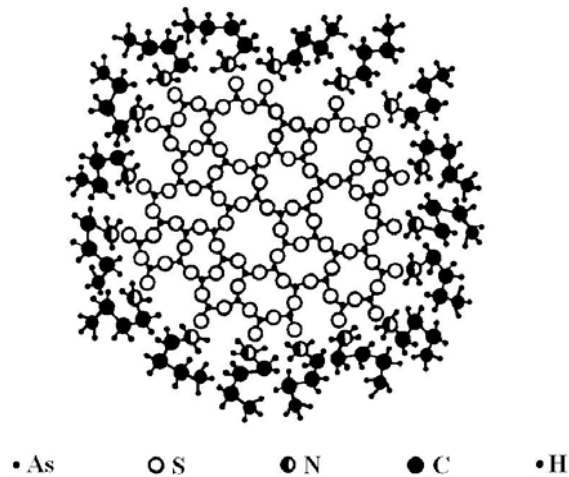

**Fig. 1:** *Schematic representation of As-S clusters (as envisaged by Chern et al. [7]) resulting from the dissolution of chalcogenide glass in amines, i.e. butylamine.*

The present work aims at elucidating the above-mentioned issue by combing results from various experimental techniques including dynamic light scattering (DLS), optical absorption spectroscopy, and AFM. The main focus is to investigate physicochemical and cluster size properties of dissolved chalcogenides glasses ($As_{33}S_{67}$ in butylamine solutions) in order to understand possible relations in morphological properties of clusters in solution and in spin-coated films. In addition, while photoinduced effects in chalcogenides glasses are exclusively reported in the glassy state, a interesting transient photoinduced effect, which takes place in solution, is revealed in the present study.

## II. EXPERIMENTAL

*1. Sample preparation*

The bulk chalcogenide glass with composition of $As_{33}S_{67}$ was prepared by mixing 5N elements of appropriate weights in evacuated quartz ampoules placed in rocking furnace (at 650 °C for 24 hours). The glass was prepared by quenching the melt to room temperature. Chalcogenide glass solutions with three different concentrations $c_1$ = 16.6, $c_2$ = 83, and $c_3$ = 166 mg/ml of $As_{33}S_{67}$ were obtained by dissolution of bulk glass in butylamine (BA) solvent (Sigma-Aldrich, 99.9%). Thin films were coated on silica glass substrates by spinning solutions with concentrations $c_1$ and $c_2$ for 30 seconds at a spin speed of ~3000 rpm. Film thicknesses were estimated of about 100 and 1700 nm, respectively. The annealing procedure that was applied to some of the spin coated films involved heating at 90 °C under moderate vacuum at 5 Pa for 1 hour.

The preparation of solutions for the DLS study was considered with care so as to avoid the presence of undesired dust particles that interfere with the scattered signal which originates from the Brownian motion of the chalcogenide glass clusters. The $As_{33}S_{67}$/BA solutions were passed through 0.2 μm (Polytetrafluoroethylene, PTFE) filters into pre-cleaned, dust-free pyrex tubing of 10 mm inner diameter. All measurements took place at ambient temperature and right angle scattering.

*2. Optical Absorption*

The electronic absorption spectra were measured with a Perkin–Elmer (model Lambda 900) spectrophotometer equipped with reverse optics and a large sample compartment. Fused silica square optical cell (Ultrasil/Helma, Germany) with an optical path length of 1.00 cm was used. The electronic absorption spectra of all solutions were recorded in the region 200 – 800 nm. The solutions were measured before and after passing through the filter in order to check for possible concentration change during filtering.

*3. Dynamic light scattering*

Normalized intensity time correlation function $g^{(2)}(q, t)=<I(q, t) I(q, 0)>/<I(q, t)>^2$, were measured over a broad time scale (from $10^{-8}$ s to $10^4$ s) using a full multiple tau digital correlator (ALV−5000/FAST) with 280 channels spaced quasi-logarithmically. The scattering wavevector $q = 4\pi n \sin(\theta/2)/\lambda_0$ depends on the scattering angle $\theta$ ($\theta = 90^o$ was used in the present work), the laser wavelength $\lambda_0$, and the refractive index of the medium *n*. Various light sources were used in order to check for possible photo-induced effects in the chalcogenide glass solutions. In particular, the following laser wavelengths were used: the 496, 488 and 514. 5 nm lines from an $Ar^+$ ion laser (Spectra Physics 2020), the 632.8 nm line from a He-Ne laser and the 671 nm line form a diode pumped solid state laser. Various power levels were used when possible. The scattered light was collected by a single mode optical fiber and transferred to a photo-multiplier and then to the digital correlator for analysis.

### 4. *Atomic force microscopy*

The AFM images of As-S film surfaces were recorded using an Atomic Force Microscopy (AFM) equipment Dimension 3100 (Digital Instruments - Veeco Metrology Group). An area of 250 x 250 nm was scanned at a high-resolution tapping mode to reveal films surface morphology. In addition an area of 5 x 5 μm was scanned in order to determine the magnitude of the root mean square roughness of the films.

### III. Results and Discussion

The optical absorption spectra for the three concentrations of $BA/As_{33}S_{67}$ glass solutions as well as for pure BA are shown in Fig. 2 in the form of transmittance. As expected, the transmission curves shift to lower energies with the increase of the glass concentration in BA. The energy gaps, $E_g$, of the solutions were estimated using the derivatives of the transmission curves. Eventually, the minimum of the first derivative corresponds to the inflection point of the transmission curve that is usually considered as an accurate estimation of $E_g$. In this way we obtained for $E_g$ the values: 2.74 eV, 2.58 eV, and 2.54 eV for $c_1$, $c_2$ and $c_3$, respectively. The concentration dependence of the energy gap is shown in the inset of Fig. 2. It is obvious that with increasing concentration – or equivalently increasing the cluster size as will become evident below – the bandgap of the solution decreases tending to that of the $As_{33}S_{67}$ bulk glass (~2.42 eV) denoted by the horizontal dashed line in the inset. The vertical arrows in Fig. 2 represent the main laser energies we employed in the present work in an effort to investigate possible photo-induced effects of the glass solutions.

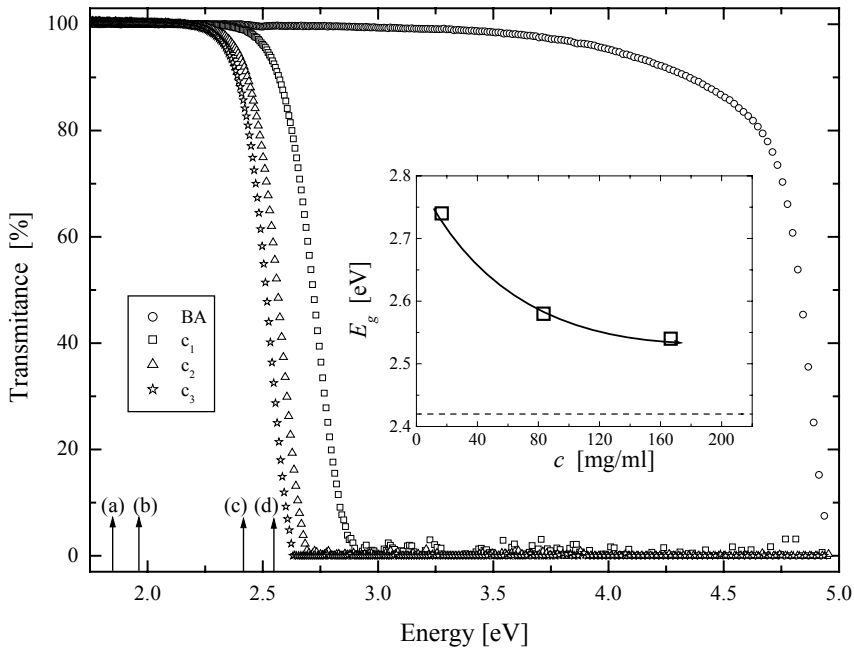

**Fig. 2:** *Transmission curves of BA and the three $BA/As_{33}S_{67}$ solutions. The arrows indicate the energies of the laser lines used in the present work, (a) 671 nm, (b) 632.8 nm, (c) 514.5 nm, and (d) 488 nm. The inset shows the concentration dependence of $E_g$ for the studied samples, the solid line is a guide to the eye and the dashed horizontal line marks the $E_g$ of the bulk glass.*

Absorption curves were recorded also for solutions before and after filtering the liquids in order to prepare the samples for DLS studies. The results revealed that there are negligible changes in the absorption features before and after filtering presumably to the capture of some tiny "insoluble" chalcogenide clusters on the filter, which however does not practically change the sample concentration.

DLS is a useful tool for studying particle sizes in solutions exploiting the Brownian motion of the suspended particles. Under the assumption of homodyne scattering conditions, which are easily fulfilled in dust-free suspensions as in the present case, the desired normalized electric-field time auto-correlation function $g^{(1)}(q,t) = \langle E(q,t) E^*(q,0) \rangle / \langle E(q,0) \rangle^2$ is related to the experimentally recorded intensity auto-correlation function $g^{(2)}(q,t) = \langle I(q,t) I(q,0) \rangle / \langle I(q,0) \rangle^2$ through the Siegert relation [11]:

$$g^{(2)}(q,t) = B\,[1 + f^* \left| g^{(1)}(q,t) \right|^2 ] \qquad (1)$$

where $B$ describes the long delay time behavior of $g^{(2)}(q, t)$ and $f^*$ represents an instrumental factor obtained experimentally from measurements of a dilute polystyrene/toluene solution. In our case, the optical fiber collection results in $f^* \cong 1$.

The electric-filed time correlation function $g^{(1)}(t)$ – for simplicity we drop the $q$-dependence in the following – was analyzed as a weighted sum of independent exponential contributions, i.e.:

$$g^{(1)}(t) = \int L(\tau) \exp(-t/\tau)\, d\tau = \int L(\ln \tau) \exp(-t/\tau)\, d\ln \tau \qquad (2)$$

where the second equality is the logarithmic representation of the relaxation times. The distribution of relaxation times $L(\ln \tau) = \tau\, L(\tau)$ was obtained by the inverse Laplace transformation (ILT) of $g^{(1)}(q, t)$ using the CONTIN algorithm [12]. The apparent hydrodynamic radii of the suspended "particles" were determined using the Stokes-Einstein relation [11],

$$R_h = \frac{k_B T}{6 \pi \eta D} \qquad (3)$$

where $k_B$ is the Boltzmann constant, $\eta$ is the viscosity of the solvent and $D$ the "particle" self-diffusion coefficient. The latter is determined by $D = 1/\tau q^2$, where $\tau$ is the relaxation time of $g^{(1)}(q,t)$.

Alternatively, $g^{(1)}(q,t)$ was fitted with a stretched exponential or Kohlrausch-Williams-Watts (KWW) function of the form:

$$g^{(1)}(t) = A \exp[-(t/\tau)^{\beta_{KWW}}], \qquad (4)$$

where $A$ is the amplitude or contrast of the time correlation function (zero-time intercept) and $\beta_{KWW}$ is the corresponding stretching exponent which is characteristic of the breadth of the distribution of the relaxation times and assumes values in the interval [0, 1]. The lower the value of $\beta_{KWW}$, the more stretched the relaxation function.

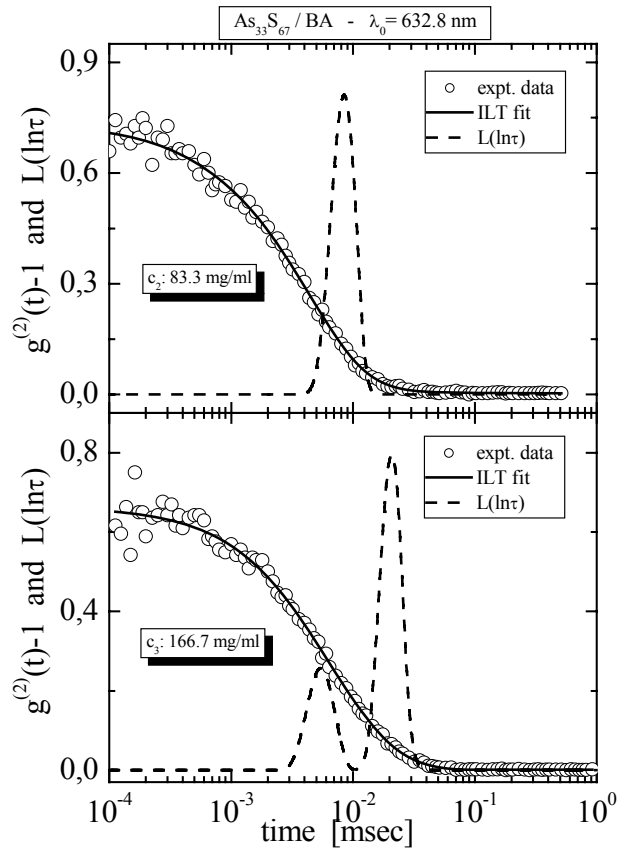

**Fig. 3:** *Intensity auto-correlation functions $g^{(2)}(t)-1$ for solutions $c_2$ and $c_3$ and their corresponding distribution of relaxation times $L(\ln \tau)$ obtained by the CONTIN analysis as described in the text.*

Most detailed measurements and analyses were preformed on the $c_2$ and $c_3$ solutions. The scattering from solution $c_1$ was very weak thus requiring long accumulation time, during which some intensity fluctuations were intervening distorting the intensity correlation function.

Nevertheless, the analysis of the relaxation data of $c_1$ (at 632.8 nm) showed that a rather good estimation of the "cluster" radius for this concentration is of about 1 nm. Figure 3 shows representative analysis examples of the solutions with concentrations $c_2$ and $c_3$. The best-fit curve (solid line through data points) was obtained with the aid of ILT (Eq. 2) and the resulting distribution of relaxation times $L(\ln \tau)$ is also shown as a dashed line. These data reveal that the sample $c_2$ is a monodispersed solution [single peak in $L(\ln \tau)$] while the denser sample ($c_3$) is polydispersed characterized by two main cluster populations. The analysis of the more dilute sample ($c_1$) studied in the present work showed that this solution is also monodispersed. These conclusions are supported by the analysis with stretched exponential functions. Indeed, the stretching exponent $\beta_{KWW}$, found after fitting with the aid of Eq. 4, is a measure of polydispersity; the lower the value of $\beta_{KWW}$ the greater the polydispersity. The magnitude of $\beta_{KWW}$ for solutions $c_2$ and $c_3$ was found to be 0.9 and 0.7, respectively. Although the difference between these two values is not drastic, the CONTIN analysis for all the accumulated time correlation functions resulted systematically in a bimodal distribution of clusters in the dense $c_3$ dispersion and a monomodal distribution of the solution $c_2$.

The diffusive nature of the cluster motion we observe in $As_{33}S_{67}$/BA solutions can be verified by inspecting the wavevector- or $q$-dependence of the diffusion coefficients. The $q$-dependence can be obtained either by varying the scattering angle or the laser wavelength, in our case the second procedure was followed. Figure 4 illustrates the $q$-dependence of the solutions with concentrations $c_2$ and $c_3$. Within experimental error the data exhibit a $q$-independent behavior as is expected for diffusive processes.

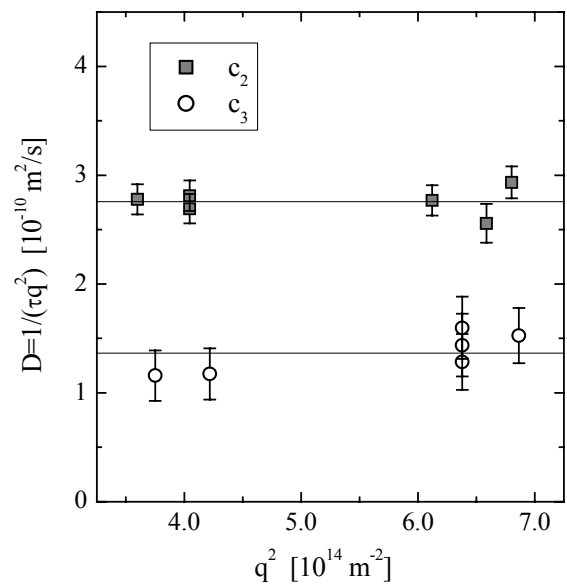

**Fig. 4:** *Wavevector dependence of the diffusion coefficient for solutions $c_2$ and $c_3$. The horizontal lines represent the mean values of D at each concentration.*

The strong dependence of the energy gap on cluster size, as revealed by optical absorption studies, calls for an investigation of a possible "spectral dependence" of the cluster size. It is well known that chalcogenide glasses exhibit a dazzling variety of photo-induced phenomena when illuminated by photons with energy near the band-gap of the glass [3-5]. It would be thus interesting to check for possible photo-induced effects in solutions containing nanoparticles of chalcogenides in solutions. Photoinduced effects have also been reported in the supercooled liquid state of corresponding glasses [5(a)], though being of dynamic origin; i.e. after the removal of the stimulus (illumination), self-annealing processes restore equilibrium.

Recording time correlation function at various laser wavelengths and various power densities on the scattering volume the following observations emerged. (i) Changes in the diffusional motion of clusters and hence on cluster size were observed only for the higher concentration ($c_3$) used in the present work. (ii) When the illumination energy was far from the energy gap of the solution $c_3$, i.e. for laser wavelengths 632.8 and 671 nm no power dependence was observed. (iii) For light energy (514.5 nm = 2.41 eV) approaching that of the band gap of $c_3$ (2.54 eV) significant photoinduced

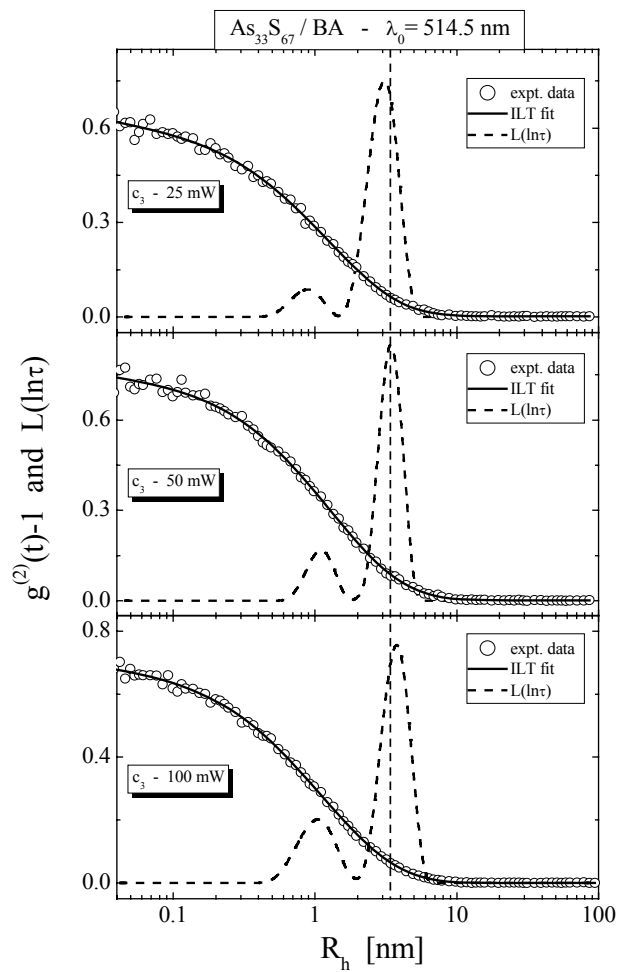

**Fig. 5:** *Intensity auto-correlation functions $g^{(2)}(t)-1$ for solution $c_3$ and their corresponding distribution of hydrodynamic radii obtained for various power levels of the incident laser beam.*

*The dashed vertical line is used as guide to reveal the shift of the cluster size distribution as a result of the athermal photo-aggregation process.*

changes are observed. In particular, a systematic *athermal photo-aggregation* phenomenon seems to take place with increasing the power density of the incident light as shown in Fig. 5. In particular, increasing the laser power by a factor of 2 and 4, the result is the increase of the hydrodynamic radius by ~11% and ~25%, respectively. These differences in hydrodynamic radii are beyond experimental error.

It could be claimed here that this is a thermal effect originating from laser radiation absorption. However, if this were true then heat induced effects would render the Brownian motion of the clusters faster, resulting in a higher diffusion coefficient of the suspended "clusters"; this should be reflected in a smaller hydrodynamic radius according to Eq. 3. Since we observe the opposite trend, i.e. increase of $R_h$ with increasing laser power, we can safely rule out that the observed effect is due to heat-induced effects. The cluster radii obtained using the procedure mentioned above are illustrated in Fig. 6. Photo-induced changes in cluster size are only observed at the highest concentration studied in this work.

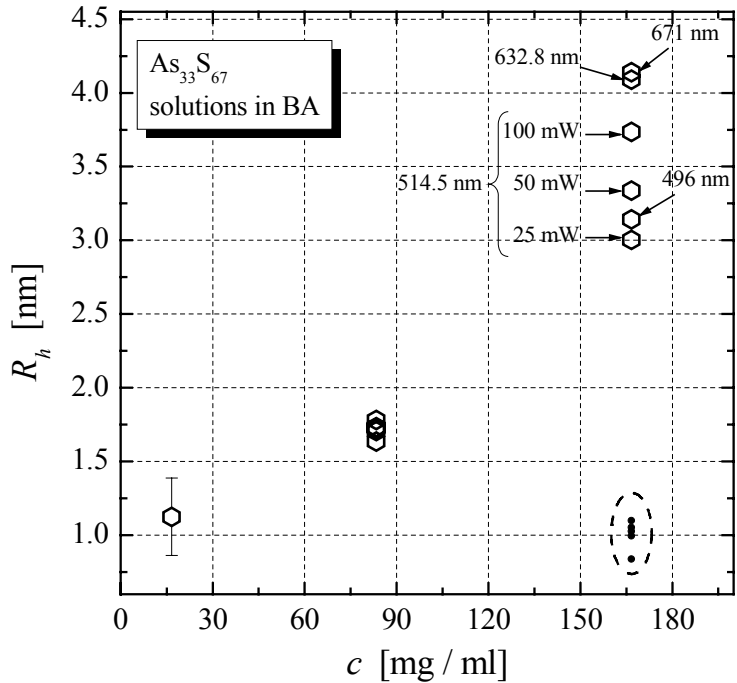

**Fig. 6:** *Spectral and concentration dependence of $As_{33}S_{67}$ glass cluster size in butylamine solutions. The cluster size grows systematical with increasing concentration. The small filled circles at the bottom right side of the figure represent the small clusters present in solution $c_3$; see text for details.*

AFM images of spin-coated As-S thin films with thicknesses 100 and 1700 nm prepared from solutions $c_1$ and $c_2$ are presented in Figs. 7(a,b) and Figs. 8(a,b), respectively. Preparation of films from the densest solution, $c_3$, was not of desired optical quality required for applications, due to high solution viscosity causing the macroscopic defects of films surface morphology. These figures illustrate the morphological details of the surface of as-deposited $As_{33}S_{67}$ films revealing the apparent grainy morphology created most likely by evaporation of the solvent during film deposition process. Film surfaces were scanned by AFM operating in the tapping mode where a tip is patting the atoms on top of the films with a specific frequency. This made it possible to reveal films local visco-elasticity by observing changes of that frequency. Such changes indicated the presence of solvent residua on surface of as-deposited thin films because solvent-rich regions exhibit higher deformability compared with solvent-poor regions.

Typical grain sizes (effective diameter) were of about 25 and 50 nm for films prepared from solutions $c_1$ and $c_2$, respectively. Evidently, a substantial smoothening of the grainy character of thin film surface morphology took place after film thermal stabilization in a vacuum oven at 90 °C and 5 Pa for 1 hour, see Figs. 7b, 8b. The evaporation of solvent during film stabilization process leads simultaneously to the increase of films roughness (at a length scale much smaller compared to the grain size) from $0.42 \pm 0.2$ nm to $2.02 \pm 0.8$ nm [9] for films prepared from $c_1$ solution and from $2.4 \pm 0.4$ nm to $5.3 \pm 0.8$ nm for films prepared from solution $c_2$. This latter effect is attributed to surface "damage" during solvent evaporation. We consider that the grainy character of surface morphology of as-deposited spin-coated thin films is related to the agglomeration of chalcogenide clusters during solvent evaporation from the film structure at the time of film production.

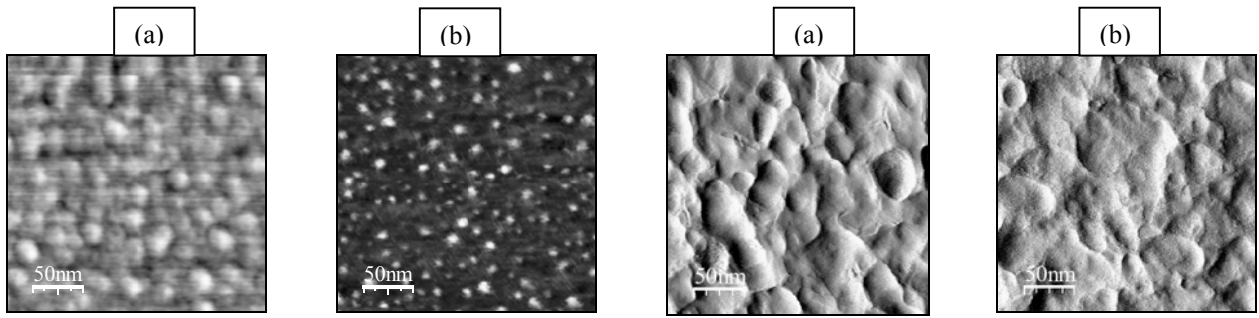

**Fig. 7:** (a) *AFM tapping mode image showing the surface morphology of as-deposited spin-coated $As_{33}S_{67}$ thin film (prepared from solution $c_1$) with thickness of 100 nm. Film surface reveals a grainy morphology as a result of the presence of solvent in thin film structure after its deposition. Typical grain sizes are of about 25 nm.* (b) *AFM tapping mode image showing the surface morphology of spin-coated $As_{33}S_{67}$ thin film with thickness of 100 nm stabilized at 90 °C for 1 hour. The grainy texture has considerably smoothed-out after solvent residua evaporation.*

**Fig. 8:** (a) *AFM tapping mode image of the surface morphology of as-deposited spin-coated $As_{33}S_{67}$ film with thickness of 1700 nm (prepared from solution $c_2$) with typical grain sizes of about 50 nm.* (b) *AFM tapping mode image of the surface morphology of 1700 nm spin-coated $As_{33}S_{67}$ film stabilized at 90 °C for 1 hour. Cluster agglomeration is apparent at elevated temperatures.*

Agglomeration of few nanometer sized chalcogenide clusters is reasonably expected in view of the concentration dependence of cluster size as revealed by DLS. In particular, during the first stages of film formation the continuously evaporating solvent leads to a progressively increasing concentration of the chalcogenide material and hence to increase of cluster size. At the end of the process (of free solvent evaporation) the cluster size, which is responsible for the grainy texture of the film surface, seems to have incorporated several hundreds of "clusters" present in the parent solution. This is apparent especially in case of films prepared from $c_2$ solution, see Figs. 8a, b. The grainy texture of film surface smoothes-out considerably after thermal stabilization because of the decrease of solvent residua concentration and the redistribution of the chalcogenide material. The removal of solvent residua causes probably the increase of films roughness as reported in [9].

## IV. CONCLUSIONS

A relatively simple method for thin film fabrication of amorphous chalcogenides involving spin-coating from proper solutions has been used to fabricate $As_{33}S_{67}$ films. In this paper we focused on the quest of a possible relationship between grain morphology of as-deposited and thermally stabilized spin-coated $As_{33}S_{67}$ chalcogenide thin films and cluster size of the glass clusters in butylamine solutions. The grainy surface pattern was found to be related to the cluster size of the dissolved chalcogenide material in the parent solution.

Optical absorption spectroscopy was used to study the optical properties of solutions at various glass concentrations. The energy gap of the dispersed chalcogenide nanoparticles was found to depend inversely proportional on concentration. Optical absorption studies enabled us to choose the proper wavelengths for performing dynamic light scattering on the solutions in order to determine the cluster size distributions at various glass concentrations. DLS data analysis revealed an almost linear dependence between cluster size and concentration. Cluster hydrodynamic radii were found between 1 and 4.2 nm for the samples studied in the present work. For the denser solution a bimodal cluster distribution was observed. A very interesting athermal photo-aggregation process was reveled were the cluster size was sensitive to the laser power in a way opposite of what was expected from thermal effects, i.e. high laser power was found to enhance cluster aggregation. Atomic force microscopy was employed to investigate the grainy texture of the surface of as-deposited and thermally stabilized spin-coated films. Typical grain sizes were of about 25 and 50 nm for the films produced by a dilute and a moderately dense solution indicating that agglomeration of clusters during solvent evaporation from the film structure, at the time of film fabrication by the spin-coating procedure, takes place. The grainy pattern was found to smooth-out after film thermal stabilization.

**Figure captions:**

**Fig. 1:** Schematic representation of As-S clusters (as envisaged by Chern *et al.* [7]) resulting from the dissolution of chalcogenide glass in amines, i.e. butylamine.

**Fig. 2:** Transmission curves of BA and the three BA/As$_{33}$S$_{67}$ solutions. The arrows indicate the energies of the laser lines used in the present work, (a) 671 nm, (b) 632.8 nm, (c) 514.5 nm, and (d) 488 nm. The inset shows the concentration dependence of $E_g$ for the studied samples, the solid line is a guide to the eye and the dashed horizontal line marks the $E_g$ of the bulk glass. Errors bars of $E_g$ values are smaller than the symbol size.

**Fig. 3:** Intensity auto-correlation functions $g^{(2)}(t)$-1 for solutions c$_2$ and c$_3$ and their corresponding distribution of relaxation times $L(\ln \tau)$ obtained by the CONTIN analysis using Eq. 2.

**Fig. 4:** Wavevector dependence of the diffusion coefficient for solutions c$_2$ and c$_3$. The horizontal lines represent the mean values of *D* at each concentration.

**Fig. 5:** Intensity auto-correlation functions $g^{(2)}(t)$-1 for solution c$_3$ and their corresponding distribution of hydrodynamic radii obtained for various power levels of the incident laser beam. The dashed vertical line is used as guide to reveal the shift of the cluster size distribution as a result of the *athermal photo-aggregation* process.

**Fig. 6:** Spectral and concentration dependence of As$_{33}$S$_{67}$ glass cluster size in butylamine solutions. The cluster size grows systematical with increasing concentration. The small filled circles at the bottom right side of the figure represent the small clusters present in solution c$_3$. Photo-induced changes in cluster size are only observed at the highest concentration studied in this work. Laser wavelengths and laser power levels for the 514.5 nm radiation are indicated in the figure. In the case of concentrations c$_2$ and c$_3$ error bars in $R_h$ are smaller that the symbol size.

**Fig. 7:** (a) AFM tapping mode image showing the surface morphology of as-deposited spin-coated As$_{33}$S$_{67}$ thin film (prepared from solution c$_1$) with thickness of 100 nm. Film surface reveals a grainy morphology as a result of the presence of solvent in thin film structure after its deposition. Typical grain sizes are of about 25 nm. (b) AFM tapping mode image showing the surface

morphology of spin-coated $As_{33}S_{67}$ thin film with thickness of 100 nm stabilized at 90 °C for 1 hour. The grainy texture has considerably smoothed-out after solvent residua evaporation.

**Fig. 8:** (a) AFM tapping mode image of the surface morphology of as-deposited spin-coated $As_{33}S_{67}$ film with thickness of 1700 nm (prepared from solution $c_2$) with typical grain sizes of about 50 nm. (b) AFM tapping mode image of the surface morphology of 1700 nm spin-coated $As_{33}S_{67}$ film stabilized at 90 °C for 1 hour. Cluster agglomeration is apparent at elevated temperatures.